\documentclass{article}

\usepackage[english]{babel}
\usepackage[super,sort,compress]{cite}

\usepackage{cite}
\usepackage{amsmath,amssymb,amsfonts}
\usepackage{algorithmic}
\usepackage{textcomp}
\usepackage{xcolor}
\usepackage{bm}
\usepackage{float}
\usepackage{subfigure}
\usepackage{graphicx}
\usepackage{float}
\usepackage{adjustbox}
\usepackage{multirow}
\usepackage{caption}
\usepackage{lineno}

\newcommand{\re}[1]{\textcolor{black}{#1}}
\newcommand{\ree}[1]{\textcolor{black}{#1}}

\newcommand{\argmin}{\arg\!\min}

\title{Kuramoto model based analysis reveals oxytocin effects on brain network dynamics}

\author{Shuhan Zheng$^1$, Zhichao Liang$^1$, Youzhi Qu$^1$,\\ Qingyuan Wu$^{3}$, Haiyan Wu$^{2,*}$, Quanying Liu$^{1,*}$}

\begin{document}

\maketitle

\begin{centering}
$^1$ Shenzhen Key Laboratory of Smart Healthcare Engineering,\\
Southern University of Science and Technology, Shenzhen 518005, China \\
$^2$ Centre for Cognitive and Brain Sciences and Department of Psychology, University of Macau, Macau, China\\
$^3$ State Key Laboratory of Cognitive Neuroscience and Learning \&
IDG/McGovern Institute for Brain Research, Beijing Normal University, Beijing 100875, China\\

$^*$ \small Corresponding to haiyanwu@um.edu.mo (H.W); liuqy@sustech.edu.cn (Q.L)
\end{centering}

\begin{abstract}
    The oxytocin effects on large-scale brain networks such as Default Mode Network (DMN) and Frontoparietal Network (FPN) have been largely studied using fMRI data. However, these studies are mainly based on the statistical correlation or Bayesian causality inference, lacking interpretability at physical and neuroscience level. 
    Here, we propose a \re{physics-based} framework of Kuramoto model to investigate oxytocin effects on the phase dynamic neural coupling in DMN and FPN.
    Testing on fMRI data of 59 participants administrated with either oxytocin or placebo, we demonstrate that oxytocin changes the topology of brain communities in DMN and FPN, leading to higher synchronization in the FPN and lower synchronization in the DMN, as well as a higher variance of the coupling strength within the DMN and more flexible coupling patterns \ree{at group level.}
    These results together indicate that oxytocin may increase the ability to overcome the corresponding internal oscillation dispersion and support the flexibility in neural synchrony in various social contexts, providing new evidence for explaining the oxytocin modulated social behaviors. Our proposed Kuramoto model-based framework can be a potential tool in network neuroscience and offers physical and neural insights into phase dynamics of the brain. 
\end{abstract}
keywords: Oxytocin Effects; Default Mode Network; Frontoparietal Network; fMRI; Kuramoto Model.

\section{Introduction}
Brain is a complex network with spatially distributed but temporally synchronized regions~\cite{sporns2011human,bassett2018nature,liu2017detecting}. The conventional methods to characterize synchronization between brain regions are mostly based on statistical properties of time series, such as Pearson correlation~\cite{samogin2019shared}, phase coherence~\cite{delorme2004eeglab} and Granger causality~\cite{wang2020large}. 
Kuramoto model, initially proposed by Japanese physicist Yoshiki Kuramoto, is a phase dynamics model to characterize the phase coupling of oscillators~\cite{kuramoto1984cooperative}. So far Kuramoto model has been extensively applied to complex network analysis, ranging from chemical networks~\cite{kuramoto2003chemical} to biological networks~\cite{bick2020understanding}.  \re{The Kuramoto model can also be used to study oscillations in the nervous system (e.g., the order parameter of the Kuramoto model reflecting the neural synchronization). Unlike statistical-based methods, the Kuramoto model can provide additional physical insights into how these oscillations are generated from internal interactions of the brain network.} 

Recently, the Kuramoto model has been introduced into network neuroscience to reveal the synchronization of neural activities across brain regions~\cite{kitzbichler2009broadband,odor2019critical}. \re{The synchronization analysis of neural dynamics at multiple scales has been applied to elucidate the existence of hierarchical modular organization in the intermediate phase of functional brain networks}~\cite{villegas2014frustrated}. Using the Kuramoto model, power-law probability distributions are found in the critical state and the human brain functional systems, which are dynamically critical in the endogenous state~\cite{kitzbichler2009broadband}.
The latest study reports the synchronization behavior of a large-scale weighted human connectome under homeostatic state. In addition, it confirms the power-law tail distribution of the time duration of this synchronization behavior in the critical exponents~\cite{odor2019critical}. 
The Kuramoto model is a convenient and effective tool for extracting the generic features of complex brain dynamics based on the parametric phase dynamics. It can reveal the strength of synchronization between brain regions through the coupling matrix, and quantify the overall coherence by the Kuramoto order parameter. The phase transition only occurs in the critical dimensions, which represents the emergence of the asynchronous phase and the synchronous phase~\cite{markram2015reconstruction}. However, despite few applications in brain networks~\cite{schmidt2015kuramoto,odor2019critical}, the potential of the Kuramoto model in neuroscience is vastly underestimated.
 
Oxytocin (OT), as a neuropeptide, can modulate neural activity and synchronization in the brain networks, \re{thereby enhancing} social adaptation and prosocial behavior~\cite{churchland2012modulating,jones2017oxytocin,ma2016oxytocin,ross2009characterization}. In the early days, researchers focused on the effect\re{s} of OT on specific brain regions, such as \re{the} amygdala~\cite{2016Intranasal} and \re{the} temporoparietal junction (TPJ)~\cite{wu2020oxytocin}. With some in-depth investigations on the brain network connectivity, more attention has turned to studying the network-level impact of OT. 
\ree{Wu et al. have found that OT significantly enhances the functional connectivity between the right TPJ and the default attention network. While the connectivity between the left TPJ and the medial prefrontal network is reduced~\cite{wu2020oxytocin}.} 
Schiller et al. has reported that OT reduces the occurrence and coverage of autonomous-processing related networks but increases the coverage of attention-related networks~\cite{2019Oxytocin}, implying that OT may reduce resources for the internal autonomous information process and allocate more resources to the external information process. The frontoparietal network (FPN) is a brain network closely related to processing external information such as language and working memory. Our previous study has found \ree{OT effects in modulating functional Pearson correlation based brain network property in the FPN and the DMN~\cite{zheng2021graph}.}

In contrast to FPN that is associated with external information processing, DMN processes more internal information~\cite{marino2019neuronal}. Many studies have shown that OT affects neural activity in DMN and its functional connectivity with other brain networks~\cite{2019Oxytocin,2020Intrinsic,wu2020oxytocin}. For example, there are reports that OT reduces the functional connectivity between DMN nodes~\cite{2017Intranasal}, reverses the interactions between DMN and the central executive network~\cite{2019Oxytocin}, and enhances the effective connectivity from the midline default network (posterior cingulate and precuneus) to the significant network~\cite{2020Intrinsic}. 
\re{The frontal network, as one of the sub-networks of DMN, serves as the upstream information processing center and receives input from the downstream area.} The study on the frontal network has found that OT strengthens the effective connectivity between brain cortex in the prefrontal and orbital prefrontal cortex~\cite{2020Intrinsic}, and changes the topography of the frontal areas and the interactions between the downstream areas such as the amygdala~\cite{Morawetz2017}.

High levels of synchronization have been found within a brain subnetwork rather than between subnetworks ~\cite{ermentrout2010mathematical}, which is also be known as enhanced within-network connectivity. \ree{It has been shown that OT can alternate several network properties in FPN~\cite{zheng2021graph},} and weaken the functional connectivity between DMN nodes~\cite{2017Intranasal}. Although OT effects on functional connectivity within specific networks are well studied, how OT affects the synchronization between nodes within/between networks is largely unknown. 
Based on previous literature, we formulate the following \re{two} hypotheses: 
\re{(1)} OT reduces the synchronization between detected nodes in DMN but enhances synchronization in FPN, which can be shown \re{by} Kuramoto-based analysis;
\re{(2)} OT increases the flexibility of DMN and FPN, which can be shown as \re{an increase in the variability} of dynamical coupling \re{modes} across time and subjects.

To test the above hypotheses, we conduct an fMRI study to investigate the effects of OT on brain network dynamics. We introduce a Kuramoto model based framework to characterize the underlying neural synchronizations in the OT group and placebo (PL) group. This framework include community detection in the network, Hilbert transform for phase information extraction, fitting of coupled modes to communities of interest in the Kuramoto model, and clustering the coupling modes. We demonstrate that the Kuramoto model, as a tool for estimating phase dynamics, has great potential to unveil the effects of OT on brain network dynamics.

\section{Method}

Here we propose a Kuramoto model-based framework to characterize neural synchronization and interactions between brain networks using fMRI data, as shown in \textbf{Figure}~\ref{Kuramotoflow}. Specifically, the framework includes the following \ree{steps}: 1) fMRI data preprocessing, 2) network detection and community detection, 3) Hilbert \re{transform on} fMRI signals to obtain phase signals, 4) fitting Kuramoto model to fMRI data with sliding window to identify coupling matrices of the communities of interest, 5) clustering analysis on the coupling matrices to obtain representative coupling patterns (or modes) across time, 6) characterize the clustered coupling modes. 

\setcounter{figure}{0}
\makeatletter 
\renewcommand{\thefigure}{\@arabic\c@figure}
\makeatother

\begin{figure*}
\centering
\includegraphics[width=0.7\textwidth]{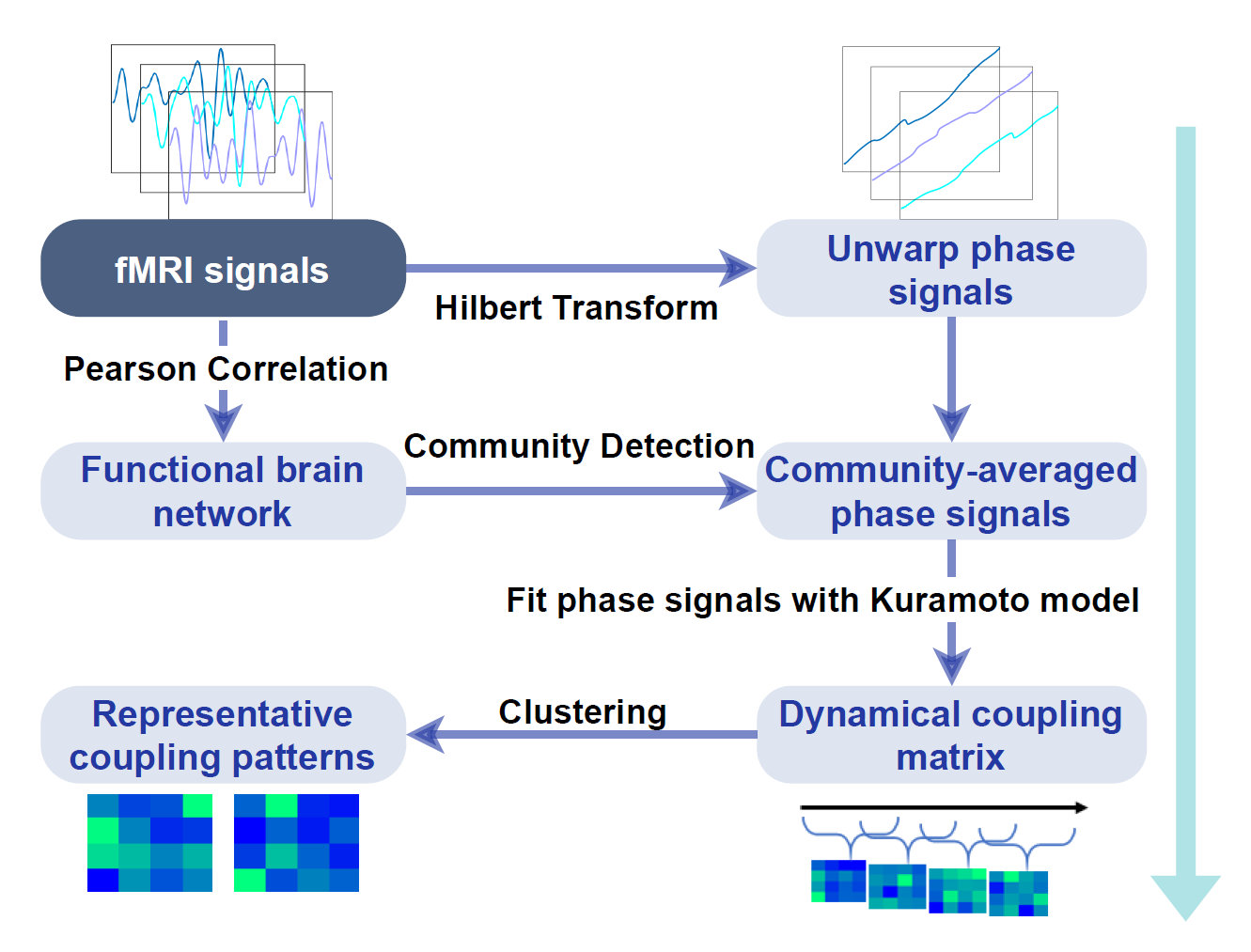}
\caption{\textbf{Kuramoto  model-based  framework} to characterize dynamical coupling in brain network. \re{The input is the preprocessed fMRI data, and the output is the detected coupling patterns.} }
\label{Kuramotoflow} 
\end{figure*}

\subsection{Kuramoto model}
The simplest form of the Kuramoto model~\cite{kuramoto2003chemical} is formulated as following: 
\begin{equation}
    \label{eq:kuramoto}
    \Dot{\theta_i}(t) = \omega_i + \frac{K}{N}\sum_{j=1}^{j=N}\Gamma(\theta_j-\theta_i)
\end{equation}
\ree{where} $\omega_i$ denotes the natural frequency of the $i^{th}$ oscillator; $K$ denotes the global coupling strength; $N$ is the number of oscillators; $\Gamma()$ represents the response function of the phase difference between two oscillators $\theta_i$ and $\theta_j$. \\

\re{In the neuron-level setting}, the function $\Gamma(\phi)$ \ree{captures} the time average phase response of voltage difference of neurons. 
Under some experimental and theoretical settings, the expression of $\Gamma(\phi)$ can be specified biologically, leading to an accurate calculation ~\cite{dodla2017effect}. In Ermentrout's book~\cite{ermentrout2010mathematical}, it has been shown that the Kuramoto model naturally arises from several assumptions of the system. These assumptions include weak coupling and asymptotic phase function approximation of the oscillator's state~\cite{ermentrout2010mathematical}. In our study, we choose $\sin(\phi)$ as the interaction function $\Gamma(\phi)$. This simplified interaction function $\sin()$ has been applied to mesoscale brain network study and reveals many critical brain functional mechanisms~\cite{kashyap2019dynamic,hellyer2014control}. The successful applications of the simplified coupling function suggests that the sinusoidal form coupling function can capture interregional dynamics. We now rewrite Eq(\ref{eq:kuramoto}) as follow,
\begin{equation}
\label{eq:kuramoto_rewrite}
    \Dot{\theta_i}(t) = \omega_i + \frac{K}{N}\sum_{j=1}^{j=N}\sin(\theta_j-\theta_i)
\end{equation}

Further, we consider heterogeneous connections \re{in the model. As we all know,} the coupling strength is not identical between interacting pairs. The coupling could be excitatory or inhibitory, depending on the value of coupling strength. So we re-write $K$ (a scalar) as $k_{ij}$ (\ree{now $K$ becomes a matrix}), and the Eq(\ref{eq:kuramoto_rewrite}) becomes:
\begin{equation}
    \label{eq:kuramoto_hetero_couple}
    \Dot{\theta_i}(t) = \omega_i + \frac{1}{N}\sum_{j=1}^{j=N}k_{ij}\sin(\theta_j-\theta_i)
\end{equation}

We use Eq(\ref{eq:kuramoto_hetero_couple}) in our study because of its simplicity and ability to capture critical dynamics, although other variants of Eq(\ref{eq:kuramoto_hetero_couple}) may provide better approximations~\cite{hellyer2014control}.

\subsection{Hilbert transform to obtain phase dynamics}

Hilbert transform has been applied in previous studies for phase analysis~\cite{kringelbach2020dynamic}. After the signals in \re{Regions of Interest (ROI)} are extracted, we perform Hilbert transform to obtain the phase signals. \ree{$X(\omega)$ and $Y(\omega)$ denote the input and output signals in the frequency domain, respectively. The Hilbert transform of $X(\omega)$ in the frequency domain can be expressed as follows:}
\begin{equation}
    \label{eq:hilbert_Y}
    Y(\omega) = X(\omega)*-\text{i}\cdot\text{sgn}(\omega)
\end{equation}
where $X(\omega)$ is the Fourier transform of $x(t)$; $sgn(\omega)$ is a sign function; $-i$ can be re-written as $e^{-i\frac{\pi}{2}}$, which is a constant. 

We expand $x(t)$ to a Fourier series \re{in time domain} \re{with $\frac{\pi}{2}$ shift on the phase of each Fourier component}. We define $y(t)$ as the output signal of Hilbert transform:
\begin{equation}
    \label{eq:hilbert_y}
    y(t) = H(x(t)) = \mathfrak{Re}(\sum_{n=-N}^{n=N}c_n*e^{i(\frac{2\pi}{T}nt-\ree{\text{sgn}(n)}{\frac{\pi}{2}})})
\end{equation}
\re{where $c_n$ is the coefficient of a Fourier component; $H(\cdot)$ represents Hilbert transform; $\mathfrak{Re}(\cdot)$ extracts the real part.} 

The phase \ree{of the} signal can be obtained by the Hilbert transformed signal $y(t)$, as $\tan(\theta) = \frac{\sin(\theta)}{\cos(\theta)}=\frac{\cos(\theta-\frac{\pi}{2})}{\cos(\theta)}$, thereby $\theta = \arctan(\frac{\cos(\theta-\frac{\pi}{2})}{\cos(\theta)})$. \re{That is, for a purely sinusoidal signal $x(t)$, the phase $\theta(t)$ can be obtained with $x(t)$ (the original signal) and $y(t)$ (the signal with $\frac{\pi}{2}$ shift on the phase) }. Since the frequency band of the filtered fMRI signal is narrow, we only obtain a small number of frequency components~\cite{gohel2015functional}. \re{In this case, although the fMRI signal is not purely sinusoidal,} the phase signal of fMRI data, $\theta(t)$, can be approximated by
\begin{equation}
\label{eq:theta}
    \theta(t) = \arctan(\frac{H(x(t))}{x(t)})
\end{equation}

\subsection{Order parameter}
In statistical mechanics, we could define an order parameter to quantify the system's behavior during phase transition. The order parameter $R(t)$ in the Kuramoto model is defined as: 
\begin{equation}
    R(t) = |\frac{1}{N}\sum_{n=1}^{n=N}e^{i\theta_n(t)}|
\end{equation}

The motivation behind the definition of $R(t)$ is intuitive. \ree{Imagine that} all oscillators are in phase, then in the complex plane, each \re{$e^{i\theta_n(t)}$} will not cancel out but direct to the same direction. In this case, $R(t)$ turns out to be 1. \re{In turn,} if all oscillators are out of phase, $R(t)$ will approximate 0. Therefore, in Kuramoto model, $R(t)$ is used to quantify the synchronization level. The higher averaged $R(t)$ indicates the higher synchronization within a system. \re{There are many other synchronization measures. Ozel et al. have compared the differences of estimating the synchronization among different measures~\cite{ozel2020intrinsic}.}

\subsection{Model fitting to estimate dynamical coupling matrix}

Here we introduce our method to fit the coupling strength $k_{ij}$ between the node $i$ and the node $j$ with the following equation:
\begin{equation}
    \Dot{\theta_i}(t) = \omega_i + \frac{1}{N}\sum_{j=1}^{j=N}k_{ij}\sin(\re{\theta_j(t)-\theta_i(t))}
\end{equation}
where the phase signal $\theta_i(t)$ and the natural frequency $\omega_i$ can be obtained by Hilbert transform and Fourier analysis, respectively. Specifically, the phase signal $\theta_i(t)$ is obtained by averaging over the detected community. \ree{Concretely, we average phase signals of nodes within each of the four communities for each subject in the OT and PL groups. $\omega_i$ is the averaged value of peak frequencies of nodes within each community.} 

By using the forward Euler method~\cite{biswas2013discussion}, we can derive the following equation,
\begin{equation}
    ( \frac{\theta_{i}(t+\Delta t)-\theta_{i}(t)}{\Delta t} - \omega_i )N = \sum_{j=1}^{j=N}k_{ij}\sin(\theta_j(t)-\theta_i(t))
\end{equation}

\re{The purpose of the fitted $\hat{k}_{ij}$ is to minimize the difference between the left and the right side.} Thus, we can estimate $\hat{k}_{ij}$ by optimizing the following loss function using pattern search.
\begin{equation}
     \hat{k}^{l}_{ij}=\argmin_{k^{l}_{ij}}||A-B(k^{l}_{ij})||_{F}^2
\end{equation}
where $||\cdot ||_{F}^2$ is the Frobenius norm; $l$ is the index of sliding window; the variable $A = (\frac{\theta_{i}(t+\Delta t)-\theta_{i}(t)}{\Delta t}-\omega_i)N$; the function $B(k^l_{ij})= \sum_{j=1}^{j=N}k^l_{ij}\sin(\theta_j(t)-\theta_i(t))$. \ree{In every sliding window, $A$ is composed of the phase signal $\theta_i$ and the signal frequency $\omega_i$. Both terms are obtained from the fMRI experiment data, so that $A$ can be calculated directly.} 

\re{To obtain the dynamic coupling term $\hat{k}_{ij}^{l}$, we apply the above calculations in every sliding window $l$ (from 1 to 6 for a subject). \ree{In every sliding window, $A$ is a matrix with 27 rows (the actual number of time points in one sliding window) and 4 columns (number of communities). $B$ has the same size as $A$.} The length of sliding window is set to be 56 seconds, and the overlapping length is 10 seconds, according to previous literature~\cite{preti2017dynamic}.}

\section{Experiments and Data Analyses}

\subsection{OT Administration and fMRI acquisition}

We recruited 59 right-handed male college students (age ranging 19 $\sim$ 26 years, and education ranging 13 $\sim$ 18 years) via an online recruiting system in Beijing. All participants provided written consent, and the research protocol was approved by the institutional review board of Beijing Normal University.
We used a double-blind placebo-controlled group design with participants randomly assigned to the oxytocin (OT) group (\re{30 participants}) or the placebo (PL) group (\re{29 participants}). 
The resting MRI scan lasted around 5 minutes, and the subjects were instructed to keep their eyes closed but not fall asleep. All images were acquired on a 3T Siemens Tim Trio scanner. More details of experiments, scanning parameters and fMRI preprocessing are shown in~\cite{wu2020oxytocin,zheng2021graph}.

\begin{figure}
\centering
\includegraphics[width=\columnwidth]{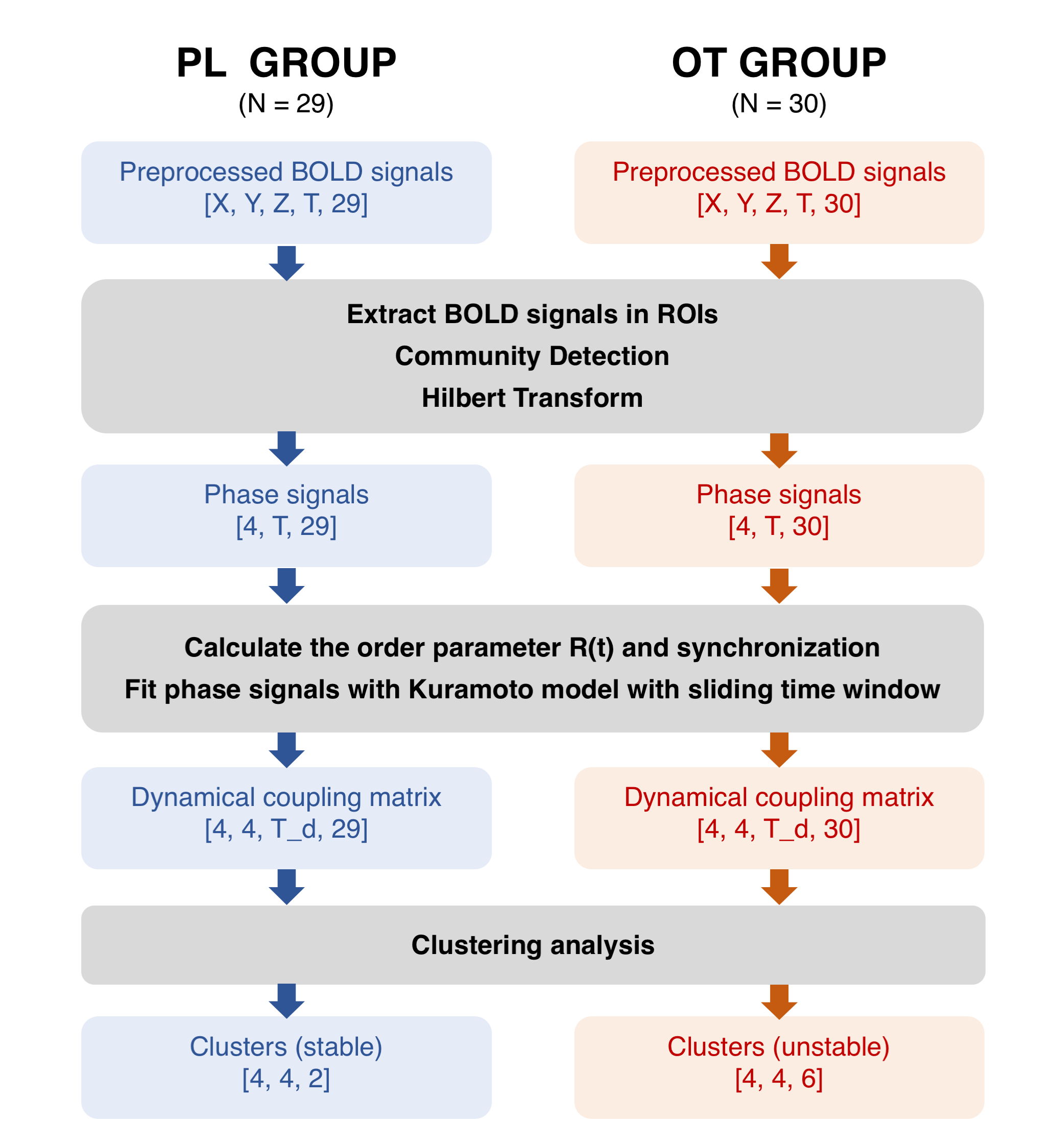}
\caption{\textbf{Data analysis pipeline}}
\label{dataflow} 
\end{figure}

\begin{table}[htbp]
\caption{\label{TableCommu_nodes}\textbf{The 4 detected communities and brain regions for OT and PL groups.}}

\begin{adjustbox}{width=\columnwidth}
\begin{tabular}{llll}
\hline
\textbf{Community} & \multicolumn{1}{l}{\textbf{Shared nodes}} & \multicolumn{1}{l}{\textbf{Nodes in OT group}} & \multicolumn{1}{l}{\textbf{Nodes in PL group}} \\ \hline

\begin{tabular}[c]{@{}c@{}}\textit{comm 1} / \textit{comm A} \\ in \textbf{FPN} \end{tabular}  & \begin{tabular}[c]{@{}l@{}}Frontal Sup R\\ Frontal Mid R\\ Frontal Mid Orb R\\ Parietal Inf R\\ Angular L\\ Angular R\\ Precuneus L\\ Precuneus R\\ Temporal Inf L\\ Temporal Inf R\end{tabular} & 
\begin{tabular}[c]{@{}l@{}}Cingulum Post L\\ Cingulum Post R \end{tabular} & 
\begin{tabular}[c]{@{}l@{}}Frontal Mid L\\ Frontal Mid Orb L\\ Parietal Inf L \end{tabular} \\ \hline

\begin{tabular}[c]{@{}c@{}}\textit{comm 2} / \textit{comm B} \\ in \textbf{FPN} \end{tabular} & 
\begin{tabular}[c]{@{}l@{}}Frontal Inf Oper L\\ Frontal Inf Oper R\\ Frontal Inf Tri L\\ Frontal Inf Tri R\\ SupraMarginal L\end{tabular} & 
\begin{tabular}[c]{@{}l@{}}Frontal Mid L \\ Frontal Inf Orb L \\ Parietal Inf L \end{tabular}& 
\begin{tabular}[c]{@{}l@{}}Precentral L\\ Cingulum Mid R \\ Parietal Sup  R\\ SupraMarginal R \end{tabular} \\ \hline

\begin{tabular}[c]{@{}c@{}}\textit{comm 3} / \textit{comm C} \\ in \textbf{DMN} \end{tabular} & 
\begin{tabular}[c]{@{}l@{}}Frontal Sup Orb L\\ Frontal Sup Orb R\\ Olfactory L\\ Olfactory R,\\ Frontal Med Orb L\\ Rectus L\\ Rectus R \\ Temporal Pole Mid L\\ Temporal Pole Mid R\end{tabular} & \begin{tabular}[c]{@{}l@{}}Amygdala L\\ Amygdala R, \\ Hippocampus L\\ Hippocampus R \\ ParaHippocampal L\\ ParaHippocampal R\end{tabular} & 
\begin{tabular}[c]{@{}l@{}}Frontal Med Orb R \\ Cingulum Post L\\ Cingulum Post R \end{tabular} \\ \hline

\begin{tabular}[c]{@{}c@{}}\textit{comm 4} / \textit{comm D} \\ in \textbf{DMN} \end{tabular} & 
\begin{tabular}[c]{@{}l@{}}Frontal Sup L\\ Frontal Sup Medial L\\ Cingulum Ant L \end{tabular} & \begin{tabular}[c]{@{}l@{}}Frontal Med Orb R\\ Cingulum Ant R \\Caudate L \\ Temporal Pole Sup L \end{tabular} & 
\begin{tabular}[c]{@{}l@{}}Frontal Inf Orb L\\ Frontal Inf Orb R\\ Frontal Sup Medial R\\ Temporal Mid R\end{tabular} \\ \hline

\end{tabular}
\end{adjustbox}
\end{table}

\subsection{Brain network community detection}
\re{In network neuroscience, segregated communities refer to dissociable cognitive components, which contain densely interconnected nodes~\cite{xu2017tri,bassett2018nature}.} We applied the community detection algorithm in both OT and PL groups. The group-level community is constructed using the virtual-typical-subject (VTS) approach~\cite{taya2016comparison}, which used group-averaged functional connectivity matrices (obtained by Pearson correlation) to extract community patterns for the PL and the OT groups. 
\re{The Louvain heuristics algorithm is a promising tool in detecting communities in the complex network~\cite{blondel2008fast}. We ran the Louvain heuristics algorithm using the Brain Connectivity Toolbox~\cite{rubinov2010complex}. It has a hyperparameter $\gamma$ to control the modular size in the community detection. A higher $\gamma$ value allows the algorithm to detect a community with a smaller size. The hyperparameter $\gamma$ was tuned to obtain the largest $Q$ value,  \ree{ that quantifies} the modularity of detected communities. \ree{A higher $Q$ value indicates a higher} modularity of network. We eventually set $\gamma = 1.9$. } Since the Louvain heuristics algorithm is an unsupervised algorithm, we iteratively ran the algorithm 100 times to obtain a \re{reliable} community structure. 

\subsection{Kuramoto model related analysis}
Kuramoto model-based phase synchronization analysis in the detected community nodes \ree{was} conducted to characterize the OT effects on neural synchronization. We firstly \ree{applied} the Hilbert transformation on fMRI data to obtain the phase signals. Then we \ree{averaged} the phase signals over each of the 4 detected communities. We further \ree{calculated} the temporal mean of the order parameter $R(t)$ in these community-averaged signals. The temporal-averaged $R(t)$ quantifies the synchronization level. To quantify the significance of 
OT effects on network synchronization, we \ree{performed} the Wilcoxon rank sum test to compare cross-community synchronizations between PL and OT groups, \ree{as well as the comparison of the coupling strength.}

To fit phase signals with the Kuramoto model, we \ree{took} a community-averaged signal as the phase of one oscillator. $N$ in Eq(\ref{eq:kuramoto_rewrite}) \ree{was} set equal to 4, corresponding to the number of oscillators. $\omega_i$ \ree{was} obtained from averaging peak frequencies (of fMRI signal) over nodes within a community. There are 6 time windows in total. We \ree{ran} our fitting algorithm for each subject in every time window (See \re{\textit{Section 2}} for details of the fitting algorithm). The details of the data analysis pipeline are summarized in \textbf{Figure}~\ref{dataflow}.

\begin{figure*}[tb]
\centering
\includegraphics[width=0.9\textwidth]{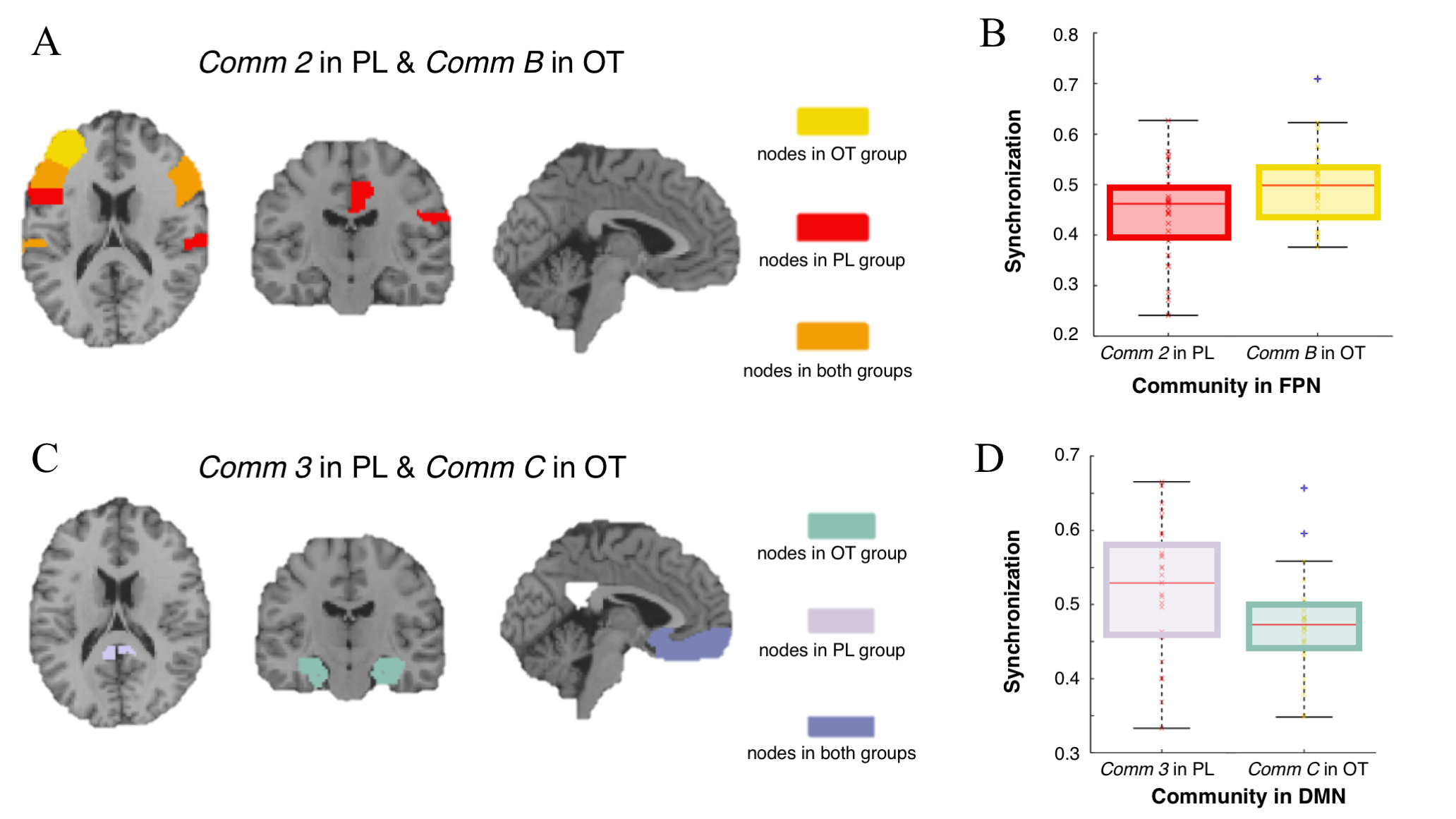}
\caption{\textbf{Synchronization analysis.} 
\textbf{(A)} Visualization of detected \ree{communities} in FPN. The detected 4 nodes in \textit{comm 2} for the PL group, 3 nodes in \textit{comm B} for the OT group and 5 shared nodes in both groups are indicated with different colors.
\textbf{(B)} Comparison of synchronization in \ree{communities} in FPN. The synchronization of the OT group is significantly higher than the PL group ($p=0.0357, z=-2.0999$).
\textbf{(C)} Visualization of detected \ree{communities} in DMN. The detected 3 nodes in \textit{comm 3} for the PL group, 6 nodes in \textit{comm C} for the OT group and 9 shared nodes in both groups are indicated with different colors.
\textbf{(D)} Comparison of synchronization in \ree{communities} in DMN. The synchronization of the OT group is significantly lower than the PL group ($p=0.0169, z=2.3880$).}
\label{figure_sync} 
\end{figure*}

\section{Results}
\subsection{OT effects on synchronization}
By community detection, we \re{discovered} 4 communities in each group. The highest $Q$ value for the PL group and the OT group are $0.1510$ and $0.1477$, respectively. The results of community detection are summarized in \textbf{Table}~\ref{TableCommu_nodes}. Our results show that both the PL and OT groups have two communities in FPN and DMN, but they involve different brain regions (see a full list of brain regions for the detected four communities in \textbf{Table}~\ref{TableCommu_nodes}).

We then further \re{compared} the synchronization level (the mean of the order parameter $R(t)$) in each community between two groups (\textbf{Figure}~\ref{figure_sync}). The statistical results show significant differences in a community related to FPN ($p=0.0357, z=-2.0999$
with \textit{ranksum} test on \textit{comm 2} in PL and \textit{comm B} in OT) and a community related to DMN ($p=0.0169, z=2.3880$ with \textit{ranksum} test on \textit{comm 3} in PL and \textit{comm C} in OT). Specifically, OT increases synchronization in \textit{comm B} in FPN but decreases \textit{comm C} in DMN. \re{The remaining two pairs of communities do not show significant difference in the synchronization level.}

\subsection{Variance of the coupling strength}

\begin{figure*}[htbp]
\centering
\includegraphics[width=0.9\textwidth]{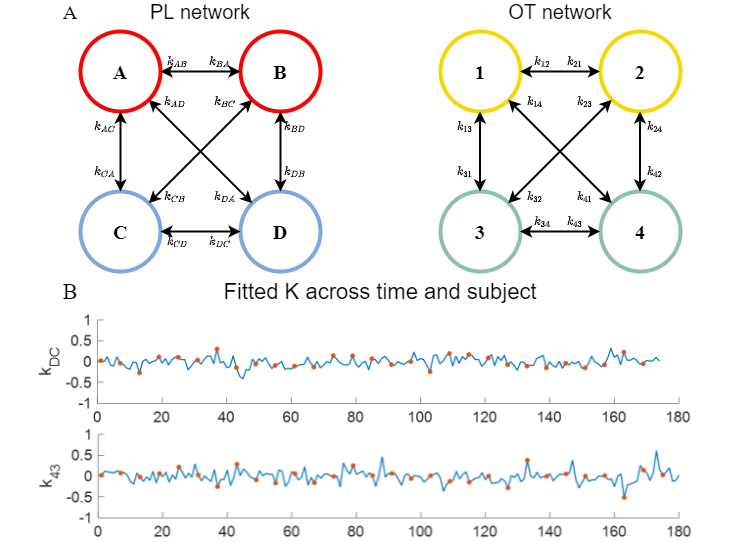}
\caption{\textbf{Cross-community coupling.} \textbf{(A)} Schematic diagram of the coupling strength: $k_{ij}$ indicates the coupling strength between two communities in the PL group (left) and the OT group (right). \re{Here, the coupling strength $k$ is directional ($k_{ij}\neq k_{ji}$). Communities in red and yellow indicate that these communities belong to FPN, while blue and dark green indicate that they belong to DMN.} \textbf{(B)} Coupling strengths across time and subjects: $k_{DC}$ (upper) and $k_{43}$ (bottom). We concatenate the dynamic $k$ of all subjects. \re{The data points between the two red dots are the dynamic $k$ values of one subject}.} 
\label{figure_fitk} 
\end{figure*}

In \textbf{Figure}~\ref{figure_fitk}A, we illustrate the naming of communities and related coupling strengths. \ree{Communities belong to FPN are colored in red and yellow, while blue and dark green communities belong to DMN.}

We calculate the variance of $k^{l}_{ij}$ for each subject. The index of time window $l$ ranges from 1 to 6. The only significant difference occurs between the variance of $k_{43}$ and $k_{DC}$ (See \textbf{Figure}~\ref{figure_fitk}B). The variance of $k_{43}$ is higher than the variance of $k_{DC}$ with $p=0.043, z=-2.0241$. The upper plot in \textbf{Figure}~\ref{figure_fitk}B corresponds to the fitted $k_{DC}$ in the PL group, and below corresponds to the $k_{43}$ in the OT group. In Figure~\ref{figure_fitk}B, the red dot separates different subjects. There are 6 data points (\re{corresponding} to 6 fitted $k$s in 6 time windows) between two neighboring red dots. Other pairs of coupling strength do not show any significant difference in the variance or the mean. \re{The interpretation of these results will be discussed in Sec~\ref{dis:variability}.}

\subsection{Clustering of dynamical coupling patterns}
Using Pearson correlation as a distance metric, we apply K-means clustering analysis on the dynamical coupling patterns across all subjects and all 6 sliding windows. In other words, there are 29*6 coupling matrices in the PL group for clustering, 30*6 coupling matrices in the OT group. 

\begin{figure*}[htbp]
\centering
\includegraphics[width=0.8\textwidth]{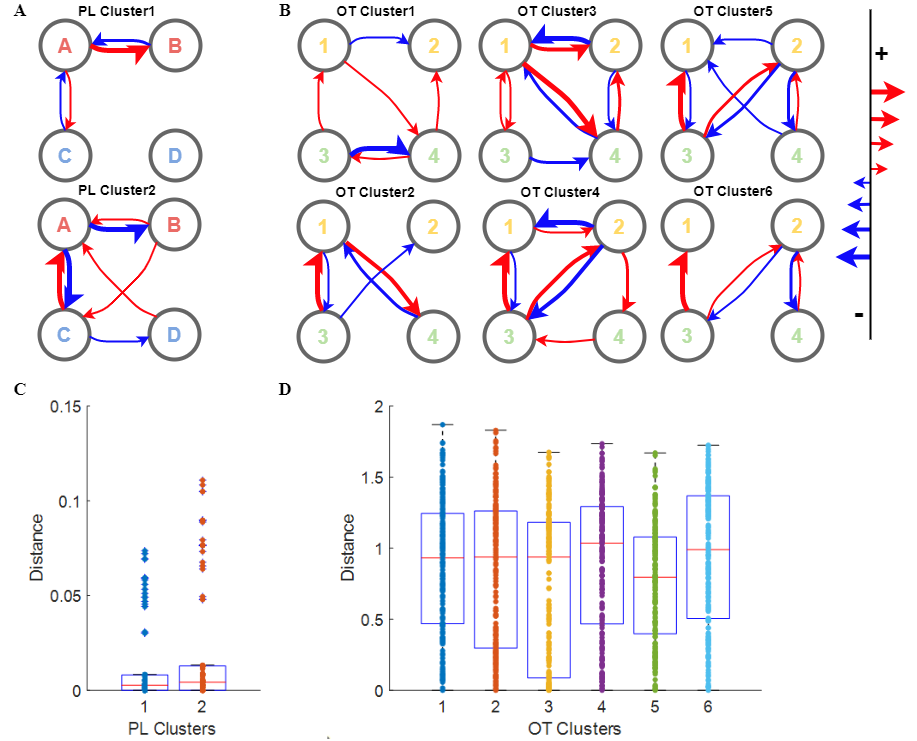}
\caption{\textbf{The detected coupling patterns in clustering analysis.} \textbf{(A)} two clusters of coupling patterns in the PL group; \re{The red and blue color indicate plus and minus sigh respectively, with the degree of arrow thickness reflects the relative value of strength}  \textbf{(B)} six clusters of coupling patterns in the OT group; \textbf{(C)} the distance of the detected clusters in the PL group with 200 iterations; \textbf{(D)} the distance of the detected clusters in the OT group with 200 iterations. The larger distance indicates a higher unstability of the coupling pattern. Our results show that the detected dynamical coupling patterns in the PL group are more stable than those in the OT group; in turn, coupling in the OT group has higher flexibility.}
\label{fig:clustering} 
\end{figure*}

By maximizing silhouette value (\re{evaluating} the performance of clustering results), we obtain the optimal number of clusters. The optimal number of clusters is 2 for the PL group and 6 for the OT group. In \textbf{Figure}~\ref{fig:clustering}, we show \re{the} obtained coupling patterns and the distance between clusters obtained from different trials, which evaluates the stability of clustering results.

In \textbf{Figure}~\ref{fig:clustering}A, we show 2 coupling patterns of corresponding averaged cluster centers. The clustering results for the PL group are stable. Through 200 iterations of the K-means algorithm (which randomly initializes the starting point), we \ree{found} small correlation-based distances between the centers obtained in the first iteration and centers obtained \ree{in the remaining} 199 iterations  (See \textbf{Figure}~\ref{fig:clustering}C). In \textbf{Figure}~\ref{fig:clustering}B, we show 6 averaged cluster centers of the OT group \ree{in one trial}. Different from the PL group, we \re{found} large distances between clusters in the OT group (See \textbf{Figure}~\ref{fig:clustering}D). \ree{The clustering results were different from trial to trial. We could not find a stable clustering pattern at the group level. In the OT group, the coupling matrices were more diverse than the PL group. }

\section{Discussion}
\subsection{Methodological Perspective}

\subsubsection{Community Detection to Identify the Functional Structure}

Recently, the identification of modular \re{topology} in the network \re{has attracted} more attention in the neuroscience community. For instance, the functional topology~\cite{crossley2013cognitive} identification can \re{distinguish} comparable groups. In this study, following our previous findings of the integration of DMN and FPN over the frontal region, we further applied the Louvain community detection algorithm to the functional connectivity matrix calculated by Pearson Correlation to identify the functional topological structure between OT and PL groups. We found that both PL and OT groups have two different communities in FPN and DMN (\textbf{Table}~\ref{TableCommu_nodes}). 
The community structure in the PL group is well in line with the nodes in DMN and FPN, while the communities from OT group do not necessarily match the standard DMN and FPN topology, although there are many overlapped regions. \re{Previous studies have identified two anticorrelated networks in slow oscillation BOLD signals~\cite{deco2011emerging,fox2005human}. The community detection approach divides the network into communities. Nodes within a community are correlated, while nodes between communities are more likely to be anticorrelated. Therefore, our detected communities are consistent with the confirmed anticorrelated subnetworks.}

As an unsupervised machine learning approach, a considerable advantage of the Louvain community detection algorithm is that it does not require any prior knowledge about community structure. The Louvain algorithm, therefore, is widely applied to detect the community structure in the brain network~\cite{garcia2018applications}. 
However, \re{in the absence of} prior knowledge, the unsupervised nature might \re{pose} challenges \re{for interpreting} the detected communities and \re{elucidating} their connections to brain function and behavior. 
Other methods that can combine prior knowledge with independent component analysis have been adopted to detect brain networks~\cite{xin2021oxytocin,2017Intranasal}, leading to robust detection of brain networks. Therefore, incorporating prior knowledge in community detection is one of the future directions of brain network science.

\subsubsection{Kuramoto Model to Estimate Phase Synchronization in fMRI Data}

It has been reported that synchronization \re{combines} sensory modalities together to produce a unified \ree{perception}\cite{ermentrout2010mathematical}. Synchronizations between brain regions~\cite{fu2020scaling,ahmadlou2012fuzzy,ahmadlou2017complexity}, as well as synchronizations between individual neurons\cite{bennett2004electrical,chiappalone2007network}, have been observed and studied in the large-scale brain network. Such synchronizations between grouped signals reflects cognitive status~\cite{ahmadlou2010wavelet}. A technical review provides a detailed introduction to synchronization in the large-scale brain network~\cite{o2018dynamics}. In this study, we used the order parameter $R(t)$ in the Kuramoto model to estimate the synchronization level in the large-scale brain network, rather than statistical dependency between signals. As mentioned in \re{ \textit{Section 2}}, the definition of $R(t)$ is straightforward. Therefore, the advantage of using $R(t)$ as the synchronization indicator roots in extraction of the phase dynamics of fMRI signals. Compared to \re{the temporal dynamics of the original fMRI signal, the phase of the signal} is normalized, facilitating comparisons between subjects and regions (to avoid baseline effects). 

The previous studies have found OT reduces the usage and duration of the autonomic processing-related microstates\re{, but favoring} the attention-related microstates\cite{schiller2019oxytocin}. A higher synchronization level means a longer functional maintenance of the corresponding network. \re{In this respect}, our findings (in Figure \ref{figure_sync}) are consistent with the OT-induced opposite effects on the autonomic processing-related microstates and the attention-related microstates.

\subsubsection{K-means Clustering Analysis to Extract Coupling Patterns}

Intuitively, the connectivity matrix $K$ reflects the collective neural activity among the brain \ree{networks}. A higher value indicates stronger coupling, which overcomes the dispersion of corresponding intrinsic frequencies to yield coherence, while a lower value denotes the asynchronous behavior\cite{breakspear2010generative,sanchez2021detecting}. 

When we calculate the variance of coupling strength for each subject among 6 time windows, it shows a significant difference of the variance between $k_{43}$ and $k_{DC}$ (Shown in \textbf{Figure \ref{figure_fitk}} B). A higher value of temporal variance indicates a more dynamical connectivity pattern. The coupling strength $k_{DC}$ in the OT group has a larger variance across time than the corresponding coupling strength $k_{43}$ in the PL group. 

Further, we perform K-means clustering on dynamical coupling strength matrices to obtain coupling patterns. The center of each cluster represents a coupling pattern. \re{Notice the symmetric nature between two coupling patterns (PL cluster 1 and PL cluster 2) in the PL group, suggesting a potential balance of interactions (Shown in Figure \ref{fig:clustering} A). For example, in the PL cluster 1, $k_{BA}$ is positive while in the PL cluster 2, $k_{AB}$ holds negative.} However, the clustering results of the OT group are not stable. The exact forms of cluster centers vary broadly from trial to trial. This variety confirms that OT may induce more flexible coupling patterns, in line with our findings in the temporal variance of coupling strength.

\subsection{Neuroscience Perspective}

\subsubsection{OT exchanged the level of synchronization in FPN and DMN}

We found that OT has a significant impact on synchronization within the resting-state networks (DMN and FPN), \ree{which supports the hypothesis (1)}. Specifically, OT reduces the synchronization within DMN \re{but} increases the synchronization within FPN. Recent evidence has showed that FPN is essential to people's ability \ree{to coordinate behavior quickly and flexibly}\cite{marek2018frontoparietal}, and these functions are more related to external information. In contrast, DMN involves the integration and refinement of existing knowledge and experience, which \ree{are internal information.} \re{The neural activation in DMN} increases when people are engaging in self-introspection~\cite{raichle2015brain,yeshurun2021default}. In this respect, the reduced synchronization within DMN and increased synchronization within FPN imply that OT group might pay more attention to others (external information) and less attention to themselves, which may explain why OT can increase people's cooperative behaviors~\cite{de2016oxytocin,yang2020oxytocin}. \ree{People will} consider more about the behavior of others and how they should interact with others, rather than pay more attention to their own benefits. Many evidences from previous studies support this view in some way~\cite{zhu2019intranasal,ne2016intranasal,pfundmair2017oxytocin,marsh2020oxytocin}; for example, OT can increase subject’s sensitivity to external cues such as aggression and pain.

\subsubsection{OT increased variability of coupling strength in DMN} 

Previous studies have shown that OT can increase the responsiveness of humans and other species in both social and non-social domains~\cite{kapetaniou2021role}, thereby improving their learning and social \re{adaptability}~\cite{olazabal2020variation,zhuang2021oxytocin}. Here we examine the variance of the coupling in the resting state networks after OT/PL administration. The increased variability \re{in OT group} confirms the \ree{hypothesis (2)} that OT increases the flexibility of brain network coupling. This finding further indicates that OT may improve the sensitivity to external cues, which may help humans and other species in social adaptation~\cite{olazabal2006species}. 
We performed a clustering analysis on the coupling patterns between the brain networks of all subjects. We \re{found a significant} effect of OT on the proliferation of coupling patterns. Specifically, OT increased the variance of coupling strength and yielded more coupling patterns among detected communities. These results may suggest that OT can increase the number of modulation strategies available to people. 

One recent study on OT shows that OT can stabilize behavior through changes, a phenomenon known as allostasis~\cite{quintana2020allostatic}. \re{They demonstrated that OT, as an allostatic hormone, regulates social and non-social behaviors by maintaining allostasis in a constantly changing environment. In order to achieve allostasis, agents must adopt different strategies in different environments. Behavioral flexibility (induced by OT) should have a corresponding flexible functional connectivity. } Our finding indicates that OT does increase instability between DMN and FPN (\textbf{Figure~\ref{fig:clustering}}), partially supporting the idea that higher instability leads to higher flexibility.

\subsection{Limitations and Future Works}
It is worth mentioning the challenges and limitations of our work. First, a \re{core} challenge in our study is to ensure that the fitted phase coupling \ree{matrix $K$} can faithfully reflect the fundamental brain dynamics. The inconsistency might happen. For example, \re{we can simulate two phase-locked oscillators using the Kuramoto model, where there is a strong coupling between these two oscillators. However, if we fit these two oscillators to the Kuramoto model}, the fitted coupling matrix $K$ may be small and cannot reflect the real system dynamics. In our study, we overcome this challenge by first applying community detection to a Pearson correlation-based functional network. One oscillator's signal is obtained by averaging phase signals over a detected community (high cohesion within an oscillator). This step guarantees that the synchronicity between our averaged-oscillators is low. Thus it can improve the authenticity of the fitted \ree{$K$}. To comprehensively study the performance of Kuramoto model in brain network analysis, it is important to compare it with other brain networks analysis methods, such as Bayesian causality inference~\cite{durante2018bayesian} and independent component analysis~\cite{zhu2021altered}\ree{; this will be made in the future, using the same data.}

More advanced brain network community detection algorithms have been proposed, which may help brain network analysis. For example, the dynamic Plex Percolation Method, with its robustness to edge noise, can capture certain stereotypical dynamic community behaviors and track dynamic community organization~\cite{martinet2020robust}. By adding prior information, the evolutionary nonnegative matrix factorization method detects a more accurate dynamic community structure~\cite{ma2017evolutionary}. \re{To identify community with multi-layer networks,} integrating nonnegative matrix factorization and topological structural information may explore more high-order information~\cite{ma2021identification,vuksanovic2019cortical}. 

Validating the Kuramoto model \re{on} multiple time scales is an interesting future direction. Brain dynamics exist on multiple time scales. Due to the time resolution limitation of fMRI signals, our studies could only reveal neural dynamics on large time scale. fMRI cannot accurately detect complex, hierarchical, and high-resolution spatiotemporal dynamics of OT effects. Other neuroimaging techniques with finer time resolution should be incorporated in our future study. \re{For instance, EEG can be an electrophysiological tool to explore the temporal stability and dynamics of resting networks~\cite{schiller2019oxytocin,liu2017detecting,fang2020dynamic}.}

There are still some unexplained neural mechanisms underlying the statistical findings of OT effects, for instance, the detected increasing or decreasing functional connectivity and the information flow among the resting-state networks. \ree{Following the previous research~\cite{schmidt2015kuramoto,vuksanovic2014functional}, we could implement simulation work to gain more understanding. Besides,} physical-related dynamic methods (e.g., Hopf Oscillators) with their interpretability may explain these physiological phenomena and drive neuroscience to a new era. 

\section{Conclusion}
\re{In the present study, we \ree{conducted} a Kuramoto based analysis of OT effects on brain network dynamics, including synchronization levels and coupling patterns. By comparing the order parameter, results indicate that OT induces higher synchronization in attention-related networks and lower synchronization in autonomous related networks. Furthermore, with sliding window fitting and k-means clustering, results show that OT leads to a\ree{n unstable} functional coupling patterns. \ree{Overall}, the proposed Kuramoto model-based analysis framework provides physical and neural insights into OT effects in the phase dynamics of the brain.  }

\section*{Acknowledgement}
This research was supported by National Natural Science Foundation of China (No. 62001205,U1736125), Guangdong Natural Science Foundation Joint Fund (No. 2019A1515111038),Guangdong natural science foundation(No.2021A1515012509), Shenzhen Key Laboratory of Smart Healthcare Engineering (ZDSYS20200811144003009).
All authors declare no conflicts of interests.

\bibliographystyle{abbrv}
\bibliography{reference.bib}

\end{document}